\documentclass[reprint,prl,superscriptaddress]{revtex4-1}
\usepackage{graphicx}
 \usepackage{amsmath}
\usepackage{bm}

\usepackage{hyperref}
\hypersetup{%
 pdftitle = {Shaping volumetric light distribution through turbid media using real-time three-dimensional optoacoustic feedback},
 pdfauthor = {H. Estrada},
 bookmarksopen = true,
 colorlinks  = true,
 citecolor   = black,
 linkcolor   = black,
 raiselinks  = true,
 urlcolor    = blue,
 pdfstartview= FitH,
}
\begin{document}

\title{Shaping volumetric light distribution through turbid media using real-time three-dimensional optoacoustic feedback}
\author{X. Lu\'is De\'an-Ben}
\thanks{These authors contributed equally to the work}
\affiliation{Institute for Biological and Medical Imaging (IBMI), Helmholtz Zentrum M\"unchen, Neuherberg, Germany}
\author{H\'ector Estrada}
\thanks{These authors contributed equally to the work}
\affiliation{Institute for Biological and Medical Imaging (IBMI), Helmholtz Zentrum M\"unchen, Neuherberg, Germany}
\author{Daniel Razansky}
\email{Corresponding author: dr@tum.de}
\affiliation{Institute for Biological and Medical Imaging (IBMI), Helmholtz Zentrum M\"unchen, Neuherberg, Germany}
\affiliation{Faculty of Medicine, Technische Universit\"at M\"unchen, Munich, Germany}

\date{\today}
\begin{abstract}
Focusing light through turbid media represents a highly fascinating challenge in modern biophotonics. The unique capability of optoacoustics for high resolution imaging of light absorption contrast in deep tissues can provide a natural and efficient feedback to control light delivery in scattering medium. While basic feasibility of using optoacoustic readings as a feedback mechanism for wavefront shaping has been recently reported, the suggested approaches may require long acquisition times making them challenging to be translated into realistic tissue environments. In an attempt to significantly accelerate dynamic wavefront shaping capabilities, we present here a feedback-based approach using real-time three-dimensional optoacoustic imaging assisted with genetic-algorithm-based optimization. The new technique offers robust performance in the presence of noisy measurements and can simultaneously control the scattered wave field in an entire volumetric region.
\end{abstract}


\maketitle 

Photons with energies within the visible and near-infrared range provide useful means for imaging and sensing the chemical and microstructural composition of matter \cite{band2006light}.  Imaging with light is however often hampered by scattering in nanoscale structures exhibiting heterogeneities in the optical refractive index \cite{mllf2012}. Of particular relevance is biological and medical imaging, where optical techniques can arguably offer the most diverse set of tools for interrogation of living tissues \cite{WeissPN2008}, yet suffer from severe performance degradation due to strong scattering at depths beyond a few hundred microns \cite{ntziachristos2010b}.
To this end, the feasibility of focusing light through strongly scattering media has been demonstrated by means of wavefront shaping techniques \cite{VelleMOL2007}, which spatially modulate the phase of the incident wavefront in order to create positive interference at a specific location in the speckle pattern of the scattered wavefield. A modification of an optical microscope using this approach has been demonstrated to overcome scattering by thin turbid layers \cite{VelleAOl2010}. However, exploiting wavefront shaping for real imaging applications implies, on the one hand, focusing light within the scattering sample and, on the other hand, probing light delivery to the desired location from outside the diffuse object. As light undergoes scattering while traveling through the object, pure optical readings can only provide low spatial resolution information \cite{ntziachristos2010b}, while the focusing capacity is further compromised due to the large number of optical modes (speckles) per unit volume \cite{PopofLFBGNJoP2011}. 
A possible solution for improving the spatial resolution is to rely upon ultrasonic resolution, which is not affected by light diffusion. One approach consists in ultrasonically tagging the light source at a given location within the scattering object and holographically recording the phase of the resulting wavefront from outside the object. This subsequently allows confining light within the focal area of the transducer by means of phase conjugation techniques \cite{xlw2011,SiFCNP2012,jwhmy2013}. Similarly, the combination between light and sound can be exploited via optoacoustic imaging \cite{wang2012}, which uses short-pulsed lasers to excite ultrasound responses within a sample by means of local absorption of light and nonradiative relaxation. As compared to other approaches, the latter method comes with important advantages for wavefront shaping as it does not require placement of additional detectors behind the imaged sample. The feasibility of using optoacoustic responses as a feedback mechanism for iteratively optimizing the brightness at a given location outside the scattering medium has been recently showcased \cite{KongSLCLCOL2011,ccdjmp2013b}. Furthermore, by measuring the complex linear relations between input and output optical modes, the optoacoustic transmission matrix approach allows more flexibility in controlling the wavefield shape in one or two dimensions \cite{ChaigKCFBGNP2014,ChaigGKBGOL2014}.
The most recent developments in optoacoustic imaging technology have enabled acquisition and rendering of three-dimensional images at a frame rate determined by the pulse repetition frequency of the laser \cite{LuisRLSA2014,DeanBRSR2014}. In this Letter, we demonstrate the feasibility of simultaneous control over volumetric light distribution through turbid media using real-time three-dimensional optoacoustic feedback.

The lay-out of the experimental system is depicted in Fig. 1(a). A frequency-doubled Q-switched Nd:YAG laser (Lab-190-30, Spectra-Physics) was used as light excitation source. The laser generates short pulses ($\sim$6~ns duration) at a wavelength of 532~nm and a pulse repetition frequency of 15~Hz. The energy per pulse was set to approximately 4~mJ. The output beam was horizontally polarized by means of a half-wave plate and a polarizing beam splitter and further collimated with a telescopic beam expander. The beam is then reflected in a spatial light modulator (SLM, PLUTO-BB II, Holoeye Photonics AG) and subsequently focused through a ground glass diffuser (Thorlabs DG10-120).
\begin{figure}[h]
 \includegraphics[width=\columnwidth]{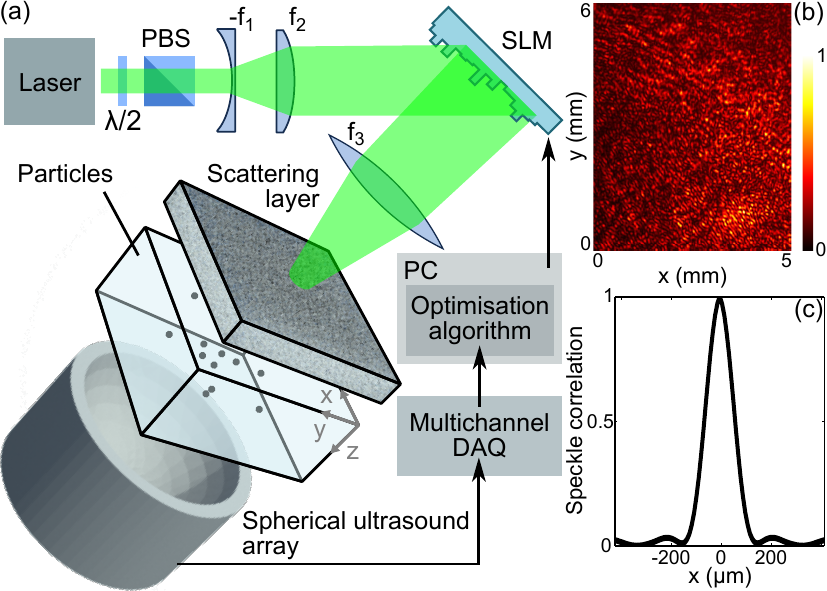}
 \caption{(a) Schematic of the experimental setup depicting the illumination arrangement, the imaged sample consisting of random distribution of absorbing microparticles, and the three-dimensional optoacoustic feedback system. (b) Image of the speckle pattern at the sample location, as captured by a CCD camera. (c) Speckle correlation function from (b)}
\end{figure}
The speckle width $w$ of the scattered beam is determined by the width of the illuminated area at the diffuser $D$ as $w=\lambda z/D$ \cite{dainty1975}
being $\lambda$ the laser wavelength and $z$ the distance from the diffuser. The speckle pattern at a distance of approximately 35 mm as imaged in a CCD sensor (IDS Imaging Development Systems GmbH, UI-2240SE-M-GL) is displayed in Fig. 1(b). The importance of the dimensions of the speckle is two-fold. On the one hand, the speckle width corresponds approximately to the width of the smallest achievable focus \cite{vlm2010} and hence determines the maximum fluence that can be delivered at this spot. On the other hand, each speckle grain represents an optical mode, and the intensity enhancement $\eta$ that can be achieved by shaping the incident wavefront is given by \cite{PopofLFBGNJoP2011}
\begin{equation}
 \eta=\frac{N_{\text{SLM}}}{2N_{\text{modes}}}\,,
\end{equation}
where $N_{\text{SLM}}$ is the number of controlled degrees of freedom (pixels in the SLM) and $N_{\text{modes}}$ is the number of optical modes contained within the volume that can be resolved in the wavefront shaping optimization procedure.
 
In our case, the feedback mechanism for the optimization is provided with a three-dimensional optoacoustic imaging system consisting of a spherical matrix array of 256 piezoelectric elements covering a solid angle of 90$^{\circ}$, as described in detail elsewhere \cite{DeanROE2013}. Each element has a central frequency of 4~MHz and -6~dB bandwidth of 100\%, thus providing nearly isotropic optoacoustic resolution of approximately  $200\;\mu\mathrm{m}$ around the center of the spherical tomographic acquisition geometry. A graphics processing unit (GPU) implementation of back-projection reconstruction allows reconstructing the optical absorption distribution in a volume of approximately 1~cm$^3$ at a faster speed than the time lapse between two laser pulses \cite{DeanORMIITo2013}, i.e. rendering three-dimensional optoacoustic images in real time.  Specifically, the value of the optoacoustic image $\text{OA}(x)$ at a point $x$ is calculated as
\begin{equation}
 \text{OA}(x)=\sum_{i}s_{p,i}(t_{i}),                          
\end{equation}
being $s_{p,i}(t_{i})$ the processed signal collected by the $i$-th transducer element. $t_{i}=d_{i}/c$ is the time-of-flight from $x$ to such transducer element, where $d_{i}$ is the propagation distance and $c$ is the speed of sound in the medium. The reconstructed three-dimensional optoacoustic image is then used to provide feedback for optimizing the wavefront shape on a per-pulse basis.
\begin{figure}[h]
 \includegraphics[width=\columnwidth]{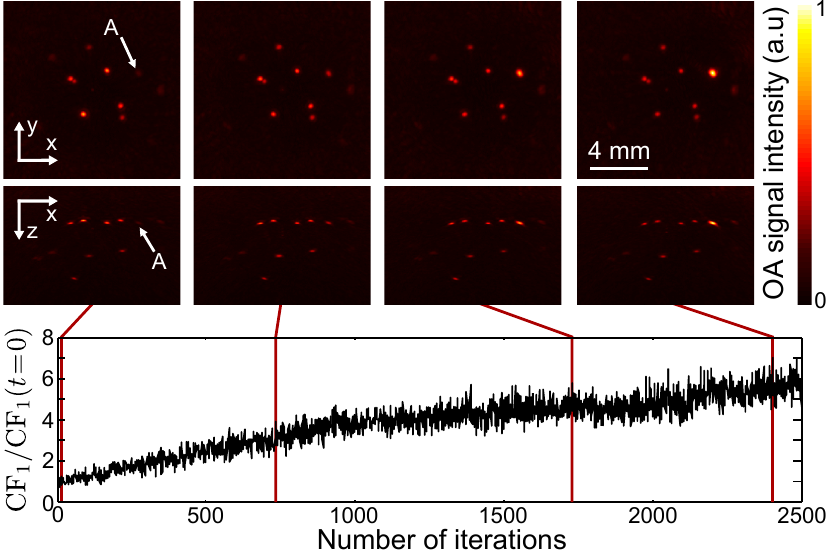}
 \caption{The time course of the wavefront shaping optimization procedure. The evolution of the cost function for all iterations is shown, along with the maximum intensity projection views of 3D optoacoustic reconstructions at 4 representative iterations of the genetic optimization algorithm. The leftmost image corresponds to a constant phase value in the SLM pixels. The particle onto which the light is focused is labeled A (see Media 1).}
\end{figure}

In order to showcase the capability of wavefront shaping with this approach, 15 mm diameter clear agar phantom was used containing sparsely distributed 200~$\mu$m-diameter absorbing black paramagnetic polyethylene microsphres (Cospheric BKPMS 180–210). The  centre of the phantom was positioned at an approximate distance of 35~mm from the diffuser, corresponding to a speckle grain size of 90~$\mu$m (Fig. 1(c)), i.e. in the range of the spatial resolution of the ultrasound array. The three-dimensional maximum intensity projections (MIP) views of the optoacoustic reconstructions of the phantom are shown in Fig. 2.  

In order to focus light on a given sphere (e.g. labeled A in the leftmost image in Fig. 2), one may consider as cost function $\text{CF}_1$ the maximum of the optoacoustic image $\text{OA}(\mathbf{r})$ in a volume of interest $\text{VOI}_{\text{A}}$ around the sphere, i.e.,
\begin{equation}
 \text{CF}_1=max\left\{\text{OA}(\text{VOI}_{\text{A}})\right\}\,.                          
\end{equation}
The value of the cost function $\text{CF}_1$ was then iteratively maximized by means of a genetic algorithm. As previously shown \cite{cbcp2012}, this particular type of algorithms is more robust to noise in comparison to other optimization techniques, such as the optoacoustic transmission matrix approach \cite{ChaigKCFBGNP2014}. The SLM pixels were grouped to form a matrix array of 20$\times$20 elements and the acquired optoacoustic signals were averaged 5 times to minimize the influence of the per-pulse energy oscillations. The time course of the optimization procedure is displayed in Fig. 2, where also evolution of the cost function for all iterations is shown. A three-dimensional rotational view of the optoacoustic reconstructions during the entire optimization procedure is provided in a movie available in the online version of the journal (see Media 1). Considering elongated speckles having a width of 90~$\mu$m, each microsphere would approximately contain 5 speckle grains (optical modes), which corresponds to theoretically achievable light intensity enhancement of approximately 40. Factors that may influence the maximum achievable enhancement include speckle decorrelation during the experiment, noise, laser energy oscillations, additional light scattering by the agar medium, and  SLM imperfections \cite{ChaigKCFBGNP2014}. In addition, the enhancement can be affected by the initial speckle distribution at the target location.

Wavefront shaping with three-dimensional feedback can be obviously exploited for more sophisticated control of the wavefield, e.g. by enhancing light intensity at several points at the same time, as demonstrated in the subsequent experiment with the same phantom. In this case, the objective was to simultaneously focus the light onto the spheres labeled B and C (Fig. 3(a)), representing the three-dimensional view of the phantom for a constant value of the phase in the SLM pixels. Two volumes of interest $\text{VOI}_{\text{B}}$ and $\text{VOI}_{\text{C}}$ around these spheres were considered with two alternative cost functions $\text{CF}_1$ and $\text{CF}_2$ defined as
\begin{equation}
  \text{CF}_2=max\{\text{OA}(\text{VOI}_{\text{B}})\} + max\{\text{OA}(\text{VOI}_{\text{C}})\}
\end{equation}
and
\begin{equation}
  \text{CF}_3=\dfrac{max\left\{\text{OA}(\text{VOI}_{\text{B}})\right\} + max\left\{\text{OA}(\text{VOI}_{\text{C}})\right\}}{\left\rvert max\left\{\text{OA}(\text{VOI}_{\text{B}})\right\} - max\left\{\text{OA}(\text{VOI}_{\text{C}})\right\}\right\rvert + k}
\end{equation}
where $k=max\left\{\text{OA}(\textbf{r})\right\}\rvert_{t=0}$.
The resulting images after 2500 iterations obtained with $\text{CF}_2$ and $\text{CF}_3$ are displayed in Figs.~3(b) and 3(c), respectively, along with the evolution of the cost function during the optimization procedure. A three-dimensional rotational view of the optoacoustic reconstructions during the optimization procedure is provided in a movie available in the online version of the journal for $\text{CF}_2$ (see Media 2) and $\text{CF}_3$ (see Media 3). At least 4 fold increase in the optoacoustic signal intensity from the two spheres is observed for both cost functions, whereas the same signal amplitude from the spheres is further achieved when performing the optimization with $\text{CF}_3$. This experiment illustrates the capability to simultaneously deliver a controlled amount of energy to specific locations in the volume by properly defining the cost function. 
\begin{figure}[h!]
 \includegraphics[width=\columnwidth]{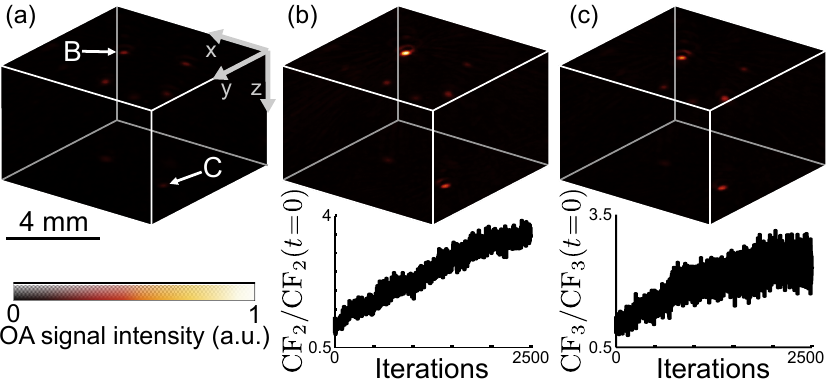}
 \caption{Three dimensional views of the first iteration (a) and the last iterations obtained with the cost functions $\text{CF}_2$ (b) and $\text{CF}_3$ (c) as defined in the text. The values of both cost functions for each iteration are displayed in (b) and (c) respectively. The particles for which light is intended to be focused are labelled as B and C. A movie version of (b) and (c) can be found in Media 2 and 3 respectively.}
\end{figure}

The results indicate the basic feasibility of controlling a scattered wavefield with optoacoustic resolution in an entire three-dimensional region. In general, as compared to pure optical approaches, optoacoustics can provide much faster feedback mechanism as our technique is based on simultaneous acquisition of information from an entire volume. More importantly, when imaging several millimeters to centimeters into turbid tissues, optoacoustic imaging can provide an order of magnitude improvement in the spatial resolution performance over optical imaging techniques based on diffuse light \cite{ntziachristos2010b}. This also implies lower number of optical modes enclosed within the resolution-limited voxel and correspondingly higher intensity enhancement with optoacoustic feedback. Yet a number of significant challenges are to be considered before the ultimate goal of light intensity enhancement within highly scattering tissues, in particular living biological tissues, is achieved. The speckle grain size inside a scattering object is expected to progressively decrease to a value in the order of $\lambda/2$. This would inevitably increase the number of optical modes that are simultaneously captured within the resolution-limited voxel of the optoacoustic probe. To overcome this, the number of degrees of freedom in the spatial light modulator must be then increased accordingly for attaining significant signal enhancement. An alternative solution to reduce the number of optical modes contained in the imaged voxel could be sought by including absorbers with smaller dimensions in the region where light needs to be focused or improving instead the spatial resolution of the optoacoustic feedback by scaling up the detected ultrasonic spectrum. An additional limitation is the coherence of the laser beam used for the wavefront shaping, which conditions the effective depth within the scattering sample for which the interference can be produced. When considering imaging in living tissues, the light focusing capacity is further challenged by motion and consequent speckle decorrelation during the optimization procedure. Fast optimization in the millisecond range is thereby required for successful wavefront shaping. Although this capacity was not achieved yet with the current configuration that uses laser pulse repetition and SLM refresh rates in the order of tens of Hz, we believe that the overall three-dimensional feedback approach introduced in this work paves the way to successful application of fast wavefront shaping techniques.

\acknowledgements{X.L.D.B. would like to acknowledge the organizers and participants of the summer school "Waves and Disorder 2014  - Ecole th\'ematique du CNRS, 1-12 July 2014, Cargese, Corsica, France" for disinterestedly sharing their deep knowledge into the subject. Valuable help from Ali Ozbek is also acknowledged. This project was supported in part by the European Research Council through the grant agreement ERC-2010-StG-260991.}


\begin{thebibliography}{22}%
\makeatletter
\providecommand \@ifxundefined [1]{%
 \@ifx{#1\undefined}
}%
\providecommand \@ifnum [1]{%
 \ifnum #1\expandafter \@firstoftwo
 \else \expandafter \@secondoftwo
 \fi
}%
\providecommand \@ifx [1]{%
 \ifx #1\expandafter \@firstoftwo
 \else \expandafter \@secondoftwo
 \fi
}%
\providecommand \natexlab [1]{#1}%
\providecommand \enquote  [1]{``#1''}%
\providecommand \bibnamefont  [1]{#1}%
\providecommand \bibfnamefont [1]{#1}%
\providecommand \citenamefont [1]{#1}%
\providecommand \href@noop [0]{\@secondoftwo}%
\providecommand \href [0]{\begingroup \@sanitize@url \@href}%
\providecommand \@href[1]{\@@startlink{#1}\@@href}%
\providecommand \@@href[1]{\endgroup#1\@@endlink}%
\providecommand \@sanitize@url [0]{\catcode `\\12\catcode `\$12\catcode
  `\&12\catcode `\#12\catcode `\^12\catcode `\_12\catcode `\%12\relax}%
\providecommand \@@startlink[1]{}%
\providecommand \@@endlink[0]{}%
\providecommand \url  [0]{\begingroup\@sanitize@url \@url }%
\providecommand \@url [1]{\endgroup\@href {#1}{\urlprefix }}%
\providecommand \urlprefix  [0]{URL }%
\providecommand \Eprint [0]{\href }%
\providecommand \doibase [0]{http://dx.doi.org/}%
\providecommand \selectlanguage [0]{\@gobble}%
\providecommand \bibinfo  [0]{\@secondoftwo}%
\providecommand \bibfield  [0]{\@secondoftwo}%
\providecommand \translation [1]{[#1]}%
\providecommand \BibitemOpen [0]{}%
\providecommand \bibitemStop [0]{}%
\providecommand \bibitemNoStop [0]{.\EOS\space}%
\providecommand \EOS [0]{\spacefactor3000\relax}%
\providecommand \BibitemShut  [1]{\csname bibitem#1\endcsname}%
\let\auto@bib@innerbib\@empty
\bibitem [{\citenamefont {Band}(2006)}]{band2006light}%
  \BibitemOpen
  \bibfield  {author} {\bibinfo {author} {\bibfnamefont {Y.}~\bibnamefont
  {Band}},\ }\href {http://books.google.com/books?id=0Rar9yGfQfgC} {\emph
  {\bibinfo {title} {Light and Matter: Electromagnetism, Optics, Spectroscopy
  and Lasers}}},\ Light and Matter\ (\bibinfo  {publisher} {John Wiley \&
  Sons},\ \bibinfo {year} {2006})\BibitemShut {NoStop}%
\bibitem [{\citenamefont {Mosk}\ \emph {et~al.}(2012)\citenamefont {Mosk},
  \citenamefont {Lagendijk}, \citenamefont {Lerosey},\ and\ \citenamefont
  {Fink}}]{mllf2012}%
  \BibitemOpen
  \bibfield  {author} {\bibinfo {author} {\bibfnamefont {A.~P.}\ \bibnamefont
  {Mosk}}, \bibinfo {author} {\bibfnamefont {A.}~\bibnamefont {Lagendijk}},
  \bibinfo {author} {\bibfnamefont {G.}~\bibnamefont {Lerosey}}, \ and\
  \bibinfo {author} {\bibfnamefont {M.}~\bibnamefont {Fink}},\ }\href
  {http://dx.doi.org/10.1038/nphoton.2012.88} {\bibfield  {journal} {\bibinfo
  {journal} {Nat Photon}\ }\textbf {\bibinfo {volume} {6}},\ \bibinfo {pages}
  {283} (\bibinfo {year} {2012})}\BibitemShut {NoStop}%
\bibitem [{\citenamefont {Weissleder}\ and\ \citenamefont
  {Pittet}(2008)}]{WeissPN2008}%
  \BibitemOpen
  \bibfield  {author} {\bibinfo {author} {\bibfnamefont {R.}~\bibnamefont
  {Weissleder}}\ and\ \bibinfo {author} {\bibfnamefont {M.~J.}\ \bibnamefont
  {Pittet}},\ }\href {http://dx.doi.org/10.1038/nature06917} {\bibfield
  {journal} {\bibinfo  {journal} {Nature}\ }\textbf {\bibinfo {volume} {452}},\
  \bibinfo {pages} {580} (\bibinfo {year} {2008})}\BibitemShut {NoStop}%
\bibitem [{\citenamefont {Ntziachristos}(2010)}]{ntziachristos2010b}%
  \BibitemOpen
  \bibfield  {author} {\bibinfo {author} {\bibfnamefont {V.}~\bibnamefont
  {Ntziachristos}},\ }\href {http://dx.doi.org/10.1038/nmeth.1483} {\bibfield
  {journal} {\bibinfo  {journal} {Nat Meth}\ }\textbf {\bibinfo {volume} {7}},\
  \bibinfo {pages} {603} (\bibinfo {year} {2010})}\BibitemShut {NoStop}%
\bibitem [{\citenamefont {Vellekoop}\ and\ \citenamefont
  {Mosk}(2007)}]{VelleMOL2007}%
  \BibitemOpen
  \bibfield  {author} {\bibinfo {author} {\bibfnamefont {I.~M.}\ \bibnamefont
  {Vellekoop}}\ and\ \bibinfo {author} {\bibfnamefont {A.~P.}\ \bibnamefont
  {Mosk}},\ }\href {\doibase 10.1364/OL.32.002309} {\bibfield  {journal}
  {\bibinfo  {journal} {Opt. Lett.}\ }\textbf {\bibinfo {volume} {32}},\
  \bibinfo {pages} {2309} (\bibinfo {year} {2007})}\BibitemShut {NoStop}%
\bibitem [{\citenamefont {Vellekoop}\ and\ \citenamefont
  {Aegerter}(2010)}]{VelleAOl2010}%
  \BibitemOpen
  \bibfield  {author} {\bibinfo {author} {\bibfnamefont {I.~M.}\ \bibnamefont
  {Vellekoop}}\ and\ \bibinfo {author} {\bibfnamefont {C.~M.}\ \bibnamefont
  {Aegerter}},\ }\href {\doibase 10.1364/OL.35.001245} {\bibfield  {journal}
  {\bibinfo  {journal} {Opt. Lett.}\ }\textbf {\bibinfo {volume} {35}},\
  \bibinfo {pages} {1245} (\bibinfo {year} {2010})}\BibitemShut {NoStop}%
\bibitem [{\citenamefont {Popoff}\ \emph {et~al.}(2011)\citenamefont {Popoff},
  \citenamefont {Lerosey}, \citenamefont {Fink}, \citenamefont {Boccara},\ and\
  \citenamefont {Gigan}}]{PopofLFBGNJoP2011}%
  \BibitemOpen
  \bibfield  {author} {\bibinfo {author} {\bibfnamefont {S.~M.}\ \bibnamefont
  {Popoff}}, \bibinfo {author} {\bibfnamefont {G.}~\bibnamefont {Lerosey}},
  \bibinfo {author} {\bibfnamefont {M.}~\bibnamefont {Fink}}, \bibinfo {author}
  {\bibfnamefont {A.~C.}\ \bibnamefont {Boccara}}, \ and\ \bibinfo {author}
  {\bibfnamefont {S.}~\bibnamefont {Gigan}},\ }\href
  {http://stacks.iop.org/1367-2630/13/i=12/a=123021} {\bibfield  {journal}
  {\bibinfo  {journal} {New Journal of Physics}\ }\textbf {\bibinfo {volume}
  {13}},\ \bibinfo {pages} {123021} (\bibinfo {year} {2011})}\BibitemShut
  {NoStop}%
\bibitem [{\citenamefont {Xu}\ \emph {et~al.}(2011)\citenamefont {Xu},
  \citenamefont {Liu},\ and\ \citenamefont {Wang}}]{xlw2011}%
  \BibitemOpen
  \bibfield  {author} {\bibinfo {author} {\bibfnamefont {X.}~\bibnamefont
  {Xu}}, \bibinfo {author} {\bibfnamefont {H.}~\bibnamefont {Liu}}, \ and\
  \bibinfo {author} {\bibfnamefont {L.~V.}\ \bibnamefont {Wang}},\ }\href
  {http://dx.doi.org/10.1038/nphoton.2010.306} {\bibfield  {journal} {\bibinfo
  {journal} {Nat Photon}\ }\textbf {\bibinfo {volume} {5}},\ \bibinfo {pages}
  {154} (\bibinfo {year} {2011})}\BibitemShut {NoStop}%
\bibitem [{\citenamefont {Si}\ \emph {et~al.}(2012)\citenamefont {Si},
  \citenamefont {Fiolka},\ and\ \citenamefont {Cui}}]{SiFCNP2012}%
  \BibitemOpen
  \bibfield  {author} {\bibinfo {author} {\bibfnamefont {K.}~\bibnamefont
  {Si}}, \bibinfo {author} {\bibfnamefont {R.}~\bibnamefont {Fiolka}}, \ and\
  \bibinfo {author} {\bibfnamefont {M.}~\bibnamefont {Cui}},\ }\href
  {http://dx.doi.org/10.1038/nphoton.2012.205} {\bibfield  {journal} {\bibinfo
  {journal} {Nat Photon}\ }\textbf {\bibinfo {volume} {6}},\ \bibinfo {pages}
  {657} (\bibinfo {year} {2012})}\BibitemShut {NoStop}%
\bibitem [{\citenamefont {Judkewitz}\ \emph {et~al.}(2013)\citenamefont
  {Judkewitz}, \citenamefont {Wang}, \citenamefont {Horstmeyer}, \citenamefont
  {Mathy},\ and\ \citenamefont {Yang}}]{jwhmy2013}%
  \BibitemOpen
  \bibfield  {author} {\bibinfo {author} {\bibfnamefont {B.}~\bibnamefont
  {Judkewitz}}, \bibinfo {author} {\bibfnamefont {Y.~M.}\ \bibnamefont {Wang}},
  \bibinfo {author} {\bibfnamefont {R.}~\bibnamefont {Horstmeyer}}, \bibinfo
  {author} {\bibfnamefont {A.}~\bibnamefont {Mathy}}, \ and\ \bibinfo {author}
  {\bibfnamefont {C.}~\bibnamefont {Yang}},\ }\href
  {http://dx.doi.org/10.1038/nphoton.2013.31} {\bibfield  {journal} {\bibinfo
  {journal} {Nat Photon}\ }\textbf {\bibinfo {volume} {7}},\ \bibinfo {pages}
  {300} (\bibinfo {year} {2013})}\BibitemShut {NoStop}%
\bibitem [{\citenamefont {Wang}\ and\ \citenamefont {Hu}(2012)}]{wang2012}%
  \BibitemOpen
  \bibfield  {author} {\bibinfo {author} {\bibfnamefont {L.~V.}\ \bibnamefont
  {Wang}}\ and\ \bibinfo {author} {\bibfnamefont {S.}~\bibnamefont {Hu}},\
  }\href {\doibase 10.1126/science.1216210} {\bibfield  {journal} {\bibinfo
  {journal} {Science}\ }\textbf {\bibinfo {volume} {335}},\ \bibinfo {pages}
  {1458} (\bibinfo {year} {2012})}\BibitemShut {NoStop}%
\bibitem [{\citenamefont {Kong}\ \emph {et~al.}(2011)\citenamefont {Kong},
  \citenamefont {Silverman}, \citenamefont {Liu}, \citenamefont {Chitnis},
  \citenamefont {Lee},\ and\ \citenamefont {Chen}}]{KongSLCLCOL2011}%
  \BibitemOpen
  \bibfield  {author} {\bibinfo {author} {\bibfnamefont {F.}~\bibnamefont
  {Kong}}, \bibinfo {author} {\bibfnamefont {R.~H.}\ \bibnamefont {Silverman}},
  \bibinfo {author} {\bibfnamefont {L.}~\bibnamefont {Liu}}, \bibinfo {author}
  {\bibfnamefont {P.~V.}\ \bibnamefont {Chitnis}}, \bibinfo {author}
  {\bibfnamefont {K.~K.}\ \bibnamefont {Lee}}, \ and\ \bibinfo {author}
  {\bibfnamefont {Y.~C.}\ \bibnamefont {Chen}},\ }\href {\doibase
  10.1364/OL.36.002053} {\bibfield  {journal} {\bibinfo  {journal} {Opt.
  Lett.}\ }\textbf {\bibinfo {volume} {36}},\ \bibinfo {pages} {2053} (\bibinfo
  {year} {2011})}\BibitemShut {NoStop}%
\bibitem [{\citenamefont {Caravaca-Aguirre}\ \emph {et~al.}(2013)\citenamefont
  {Caravaca-Aguirre}, \citenamefont {Conkey}, \citenamefont {Dove},
  \citenamefont {Ju}, \citenamefont {Murray},\ and\ \citenamefont
  {Piestun}}]{ccdjmp2013b}%
  \BibitemOpen
  \bibfield  {author} {\bibinfo {author} {\bibfnamefont {A.~M.}\ \bibnamefont
  {Caravaca-Aguirre}}, \bibinfo {author} {\bibfnamefont {D.~B.}\ \bibnamefont
  {Conkey}}, \bibinfo {author} {\bibfnamefont {J.~D.}\ \bibnamefont {Dove}},
  \bibinfo {author} {\bibfnamefont {H.}~\bibnamefont {Ju}}, \bibinfo {author}
  {\bibfnamefont {T.~W.}\ \bibnamefont {Murray}}, \ and\ \bibinfo {author}
  {\bibfnamefont {R.}~\bibnamefont {Piestun}},\ }\href {\doibase
  10.1364/OE.21.026671} {\bibfield  {journal} {\bibinfo  {journal} {Opt.
  Express}\ }\textbf {\bibinfo {volume} {21}},\ \bibinfo {pages} {26671}
  (\bibinfo {year} {2013})}\BibitemShut {NoStop}%
\bibitem [{\citenamefont {Chaigne}\ \emph
  {et~al.}(2014{\natexlab{a}})\citenamefont {Chaigne}, \citenamefont {Katz},
  \citenamefont {Boccara}, \citenamefont {Fink}, \citenamefont {Bossy},\ and\
  \citenamefont {Gigan}}]{ChaigKCFBGNP2014}%
  \BibitemOpen
  \bibfield  {author} {\bibinfo {author} {\bibfnamefont {T.}~\bibnamefont
  {Chaigne}}, \bibinfo {author} {\bibfnamefont {O.}~\bibnamefont {Katz}},
  \bibinfo {author} {\bibfnamefont {A.}~\bibnamefont {Boccara}}, \bibinfo
  {author} {\bibfnamefont {M.}~\bibnamefont {Fink}}, \bibinfo {author}
  {\bibfnamefont {E.}~\bibnamefont {Bossy}}, \ and\ \bibinfo {author}
  {\bibfnamefont {S.}~\bibnamefont {Gigan}},\ }\href
  {http://dx.doi.org/10.1038/nphoton.2013.307} {\bibfield  {journal} {\bibinfo
  {journal} {Nat Photon}\ }\textbf {\bibinfo {volume} {8}},\ \bibinfo {pages}
  {58} (\bibinfo {year} {2014}{\natexlab{a}})}\BibitemShut {NoStop}%
\bibitem [{\citenamefont {Chaigne}\ \emph
  {et~al.}(2014{\natexlab{b}})\citenamefont {Chaigne}, \citenamefont {Gateau},
  \citenamefont {Katz}, \citenamefont {Bossy},\ and\ \citenamefont
  {Gigan}}]{ChaigGKBGOL2014}%
  \BibitemOpen
  \bibfield  {author} {\bibinfo {author} {\bibfnamefont {T.}~\bibnamefont
  {Chaigne}}, \bibinfo {author} {\bibfnamefont {J.}~\bibnamefont {Gateau}},
  \bibinfo {author} {\bibfnamefont {O.}~\bibnamefont {Katz}}, \bibinfo {author}
  {\bibfnamefont {E.}~\bibnamefont {Bossy}}, \ and\ \bibinfo {author}
  {\bibfnamefont {S.}~\bibnamefont {Gigan}},\ }\href {\doibase
  10.1364/OL.39.002664} {\bibfield  {journal} {\bibinfo  {journal} {Opt.
  Lett.}\ }\textbf {\bibinfo {volume} {39}},\ \bibinfo {pages} {2664} (\bibinfo
  {year} {2014}{\natexlab{b}})}\BibitemShut {NoStop}%
\bibitem [{\citenamefont {Luis Dean-Ben}\ and\ \citenamefont
  {Razansky}(2014)}]{LuisRLSA2014}%
  \BibitemOpen
  \bibfield  {author} {\bibinfo {author} {\bibfnamefont {X.}~\bibnamefont {Luis
  Dean-Ben}}\ and\ \bibinfo {author} {\bibfnamefont {D.}~\bibnamefont
  {Razansky}},\ }\href {http://dx.doi.org/10.1038/lsa.2014.18} {\bibfield
  {journal} {\bibinfo  {journal} {Light Sci Appl}\ }\textbf {\bibinfo {volume}
  {3}},\ \bibinfo {pages} {e137} (\bibinfo {year} {2014})}\BibitemShut
  {NoStop}%
\bibitem [{\citenamefont {De\'an-Ben}\ \emph {et~al.}(2014)\citenamefont
  {De\'an-Ben}, \citenamefont {Bay},\ and\ \citenamefont
  {Razansky}}]{DeanBRSR2014}%
  \BibitemOpen
  \bibfield  {author} {\bibinfo {author} {\bibfnamefont {X.~L.}\ \bibnamefont
  {De\'an-Ben}}, \bibinfo {author} {\bibfnamefont {E.}~\bibnamefont {Bay}}, \
  and\ \bibinfo {author} {\bibfnamefont {D.}~\bibnamefont {Razansky}},\ }\href
  {http://dx.doi.org/10.1038/srep05878} {\bibfield  {journal} {\bibinfo
  {journal} {Sci. Rep.}\ }\textbf {\bibinfo {volume} {4}},\  (\bibinfo {year}
  {2014})}\BibitemShut {NoStop}%
\bibitem [{\citenamefont {Dainty}(1984)}]{dainty1975}%
  \BibitemOpen
  \bibfield  {author} {\bibinfo {author} {\bibfnamefont {J.~C.}\ \bibnamefont
  {Dainty}},\ }\href@noop {} {\emph {\bibinfo {title} {Laser Speckle and
  Related Phenomena}}},\ \bibinfo {edition} {2nd}\ ed.,\ edited by\ \bibinfo
  {editor} {\bibfnamefont {J.~C.}\ \bibnamefont {Dainty}}\ (\bibinfo
  {publisher} {Springer Verlag, Heidelberg},\ \bibinfo {year}
  {1984})\BibitemShut {NoStop}%
\bibitem [{\citenamefont {Vellekoop}\ \emph {et~al.}(2010)\citenamefont
  {Vellekoop}, \citenamefont {Lagendijk},\ and\ \citenamefont
  {Mosk}}]{vlm2010}%
  \BibitemOpen
  \bibfield  {author} {\bibinfo {author} {\bibfnamefont {I.~M.}\ \bibnamefont
  {Vellekoop}}, \bibinfo {author} {\bibfnamefont {A.}~\bibnamefont
  {Lagendijk}}, \ and\ \bibinfo {author} {\bibfnamefont {A.~P.}\ \bibnamefont
  {Mosk}},\ }\href {http://dx.doi.org/10.1038/nphoton.2010.3} {\bibfield
  {journal} {\bibinfo  {journal} {Nat Photon}\ }\textbf {\bibinfo {volume}
  {4}},\ \bibinfo {pages} {320} (\bibinfo {year} {2010})}\BibitemShut {NoStop}%
\bibitem [{\citenamefont {De\'{a}n-Ben}\ and\ \citenamefont
  {Razansky}(2013)}]{DeanROE2013}%
  \BibitemOpen
  \bibfield  {author} {\bibinfo {author} {\bibfnamefont {X.~L.}\ \bibnamefont
  {De\'{a}n-Ben}}\ and\ \bibinfo {author} {\bibfnamefont {D.}~\bibnamefont
  {Razansky}},\ }\href {\doibase 10.1364/OE.21.028062} {\bibfield  {journal}
  {\bibinfo  {journal} {Opt. Express}\ }\textbf {\bibinfo {volume} {21}},\
  \bibinfo {pages} {28062} (\bibinfo {year} {2013})}\BibitemShut {NoStop}%
\bibitem [{\citenamefont {Dean-Ben}\ \emph {et~al.}(2013)\citenamefont
  {Dean-Ben}, \citenamefont {Ozbek},\ and\ \citenamefont
  {Razansky}}]{DeanORMIITo2013}%
  \BibitemOpen
  \bibfield  {author} {\bibinfo {author} {\bibfnamefont {X.}~\bibnamefont
  {Dean-Ben}}, \bibinfo {author} {\bibfnamefont {A.}~\bibnamefont {Ozbek}}, \
  and\ \bibinfo {author} {\bibfnamefont {D.}~\bibnamefont {Razansky}},\ }\href
  {\doibase 10.1109/TMI.2013.2272079} {\bibfield  {journal} {\bibinfo
  {journal} {Medical Imaging, IEEE Transactions on}\ }\textbf {\bibinfo
  {volume} {32}},\ \bibinfo {pages} {2050} (\bibinfo {year}
  {2013})}\BibitemShut {NoStop}%
\bibitem [{\citenamefont {Conkey}\ \emph {et~al.}(2012)\citenamefont {Conkey},
  \citenamefont {Brown}, \citenamefont {Caravaca-Aguirre},\ and\ \citenamefont
  {Piestun}}]{cbcp2012}%
  \BibitemOpen
  \bibfield  {author} {\bibinfo {author} {\bibfnamefont {D.~B.}\ \bibnamefont
  {Conkey}}, \bibinfo {author} {\bibfnamefont {A.~N.}\ \bibnamefont {Brown}},
  \bibinfo {author} {\bibfnamefont {A.~M.}\ \bibnamefont {Caravaca-Aguirre}}, \
  and\ \bibinfo {author} {\bibfnamefont {R.}~\bibnamefont {Piestun}},\ }\href
  {\doibase 10.1364/OE.20.004840} {\bibfield  {journal} {\bibinfo  {journal}
  {Opt. Express}\ }\textbf {\bibinfo {volume} {20}},\ \bibinfo {pages} {4840}
  (\bibinfo {year} {2012})}\BibitemShut {NoStop}%
\end{thebibliography}

%

\end{document}